\documentstyle[prl,twocolumn,aps,amssymb,psfig,latexsym]{revtex}

\begin{document}

\draft

\twocolumn[\hsize\textwidth\columnwidth\hsize\csname
@twocolumnfalse\endcsname
\title{Quantum computing with spin cluster qubits}
\author{Florian Meier$^1$, Jeremy Levy$^{2,3}$, and
Daniel Loss$^{1,3}$}
\address{$^1$Department of Physics and Astronomy, University of Basel, 
Klingelbergstrasse 82, 4056 Basel, Switzerland \newline
$^2$Department of Physics and Astronomy, University of Pittsburgh, 
Pittsburgh, Pennsylvania 15260 \newline
$^3$Center for Oxide-Semiconductor Materials for Quantum Computation,
Pittsburgh, Pennsylvania 15260
}
\date{June 17, 2002}

\maketitle

\begin{abstract}
We study the low energy states of finite spin chains with isotropic 
(Heisenberg)
and anisotropic (XY and Ising--like) antiferromagnetic 
exchange interaction with uniform
and non--uniform coupling constants. 
We show that for an odd
number of sites a spin cluster qubit can be defined in terms of 
the ground state doublet. This qubit is remarkably 
insensitive to the placement and coupling anisotropy of spins 
within the cluster.  One-- and 
two--qubit quantum gates can be generated by magnetic fields and
inter--cluster exchange, and 
leakage during quantum gate operation is small. Spin cluster qubits 
inherit the long decoherence times and short gate operation times 
of single spins. Control of single 
spins is hence not necessary for the realization of universal quantum gates.
\end{abstract}

\pacs{03.67.Lx, 75.10.Jm}
\vskip2pc]

Quantum computers outperform classical computers on certain 
tasks~\cite{bennett:00,shor:94,ekert:96,grover:97}.
The main challenge on the way to a universal quantum computer
is to achieve control over single quantum mechanical 
two state systems (qubits) while preserving long decoherence times. 
Electron~\cite{loss:98,burkard:99} and nuclear~\cite{privman:98,kane:98} 
spins have been identified 
as promising candidates for qubits because they are natural
two state systems and decoherence times for the spin degree of
freedom are unusually large~\cite{kikkawa:97,feher:59}.

For both electron~\cite{loss:98} and nuclear spin~\cite{kane:98} qubits,
one--qubit gates can be realized by local magnetic fields or by electrically  
tuning a single
spin into resonance with an oscillating field. Two--qubit
gates rely on electrical control of the exchange interaction between 
neighboring electron spins. However, even for electrons in quantum
dots with a typical diameter of $50$ nm, the required local control over 
electrical and magnetic fields is challenging. One possibility
to circumvent the problem of either local magnetic 
fields~\cite{divincenzo:00} or local exchange 
interaction~\cite{benjamin:02} is to encode the qubit in several spins. 
Such encoding has
also been studied in the context of coherence--preserving 
qubits~\cite{bacon:01}. However, all these schemes still require control
at the single--spin level.

\begin{figure}
\centerline{\psfig{file=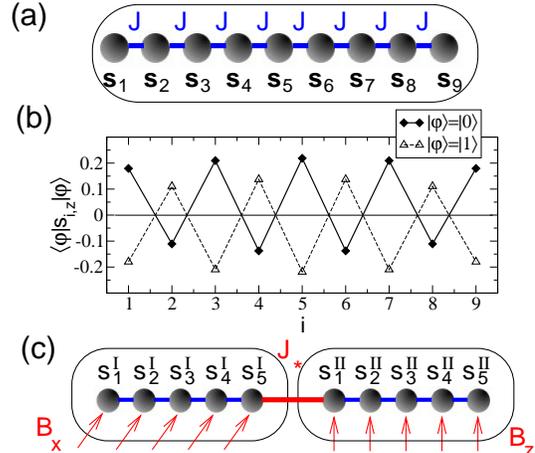,width=7cm}}
\caption{
(a) The states of the spin cluster
[Eq.~(\ref{eq:anchain})] define the spin cluster qubit. 
(b) $|0\rangle$ and $|1\rangle$ have a complicated representation
in the single--spin product basis, as evidenced by the local spin density.
(c) Quantum gates are generated by magnetic fields or $g$--factor engineering
(one--qubit gates) and a switchable inter--qubit exchange $J_\ast(t)$ 
(two--qubit gates).
}
\label{Fig0}
\end{figure}

More generally, the requirements on both local magnetic and electrical
fields can be relaxed by 
increasing the size of the qubit.
In the present work, we show that, for a wide class of antiferromagnetic spin
$s=1/2$ chains with an odd number of sites, $n_c$,
\begin{equation}
\hat{H} 
= \sum_{i=1}^{n_c-1}f_i [J_\perp  (\hat{s}_{i,x} \hat{s}_{i+1,x} + 
\hat{s}_{i,y} \hat{s}_{i+1,y}) + J_z \hat{s}_{i,z} \hat{s}_{i+1,z}],
\label{eq:anchain}
\end{equation}
the ground state doublet of the array [Fig.~\ref{Fig0}(a)] can define a new 
``spin cluster qubit''  for which quantum
gate operation times and decoherence rates increase only moderately with
array size. These features are surprisingly stable with respect
to anisotropy ($J_\perp \neq J_z$) and spatial variation (described by $f_i$) 
of the intracluster exchange, the  
spatial shape of the fields controlling quantum gate operation, and the cluster
dimension. Spin cluster qubits can be realized in a wide variety of systems,
e.g. arrays of quantum  dots~\cite{loss:98,levy:01}, 
clusters of P atoms in a Si 
matrix~\cite{kane:98}, and electron spins in molecular magnets. In contrast
to the encoded qubits suggested in earlier 
work~\cite{divincenzo:00,benjamin:02,bacon:01}, quantum computation with
spin cluster qubits is possible without control over local
spin interactions. 

{\it Isotropic spin chains as qubits. --} 
For illustration, we first discuss a spin chain with
isotropic uniform exchange, $J_\perp = J_z > 0$ and $f_i  \equiv 1$ in 
Eq.~(\ref{eq:anchain}). 
Energy eigenstates can be 
labeled according to their quantum numbers of total spin $\hat{\bf S}=
\sum_{i=1}^{n_c} \hat{\bf s}_{i}$ and the $z$--component of total spin, 
$\hat{S}_z$.
Because of the antiferromagnetic exchange, states in 
which the total spin of the chain is minimized are energetically most 
favorable~\cite{lieb:62}.
For odd $n_c$, the 
ground state is a $S=1/2$ doublet separated from the next excited 
state by a gap 
 $\Delta \sim  J\pi^2/2n_c$ determined by the lower bound of the
des Cloiseaux-Pearson spectrum. We define the spin cluster qubit in terms of 
the $S=1/2$ ground state
doublet by $\hat{S}_z|0\rangle = (\hbar/2) |0\rangle$ and
$\hat{S}_z|1\rangle = -(\hbar/2) |1\rangle$. The states 
$\{|0\rangle,|1\rangle \}$ do not, in
general, have a simple representation in the single spin product basis, but
rather are  
superpositions of $n_c!/[(n_c-1)/2]![(n_c+1)/2]!$ states [Fig.~\ref{Fig0}(b)]. 
For example, for the simplest nontrivial
spin cluster qubit with $n_c=3$,
\begin{eqnarray}
|0\rangle &=& \frac{2}{\sqrt{6}}|\uparrow\rangle_1|
\downarrow\rangle_2 |\uparrow \rangle_{3} -
\frac{1}{\sqrt{6}}|\uparrow\rangle_1|
\uparrow\rangle_2 |\downarrow \rangle_{3}  \nonumber \\ && \hspace*{1cm}
- \frac{1}{\sqrt{6}}|\downarrow\rangle_1|
\uparrow\rangle_2 |\uparrow \rangle_{3}, \label{eq:qubit-ex}
\end{eqnarray}
and $|1\rangle$ is obtained by flipping all spins.

In spite of their complicated representation in the single--spin product 
basis, 
$|0\rangle$ and $|1\rangle$ are in many respects very similar 
to the states $|\uparrow\rangle$ and $|\downarrow\rangle$ of a single spin
and, hence, can be used as qubit 
states for universal quantum computing~\cite{loss:98}. 
Because $\{|0\rangle,|1\rangle \}$ belong to one 
$S=1/2$ doublet  
such that $\hat{S}^-|0\rangle = \hbar |1\rangle$, and 
$\hat{S}^+|1\rangle = \hbar |0\rangle$ where $\hat{S}^{\pm}
= \hat{S}_x \pm i \hat{S}_y$, a 
magnetic field ${\bf B}$ constant over the cluster  
acts on the spin cluster qubit in the same way as on a single-spin qubit.
Hence, both the one--qubit phase shift and the one--qubit rotation gate can
be generated by magnetic fields $B_z(t)$ and $B_x(t)$, respectively, possibly
in combination with $g$--factor engineering~\cite{loss:98,salis:01}.
For a given $B_{z,x}(t)$, operation times
of one--qubit gates are equal to the ones for the 
single-spin qubit. We note that, due to quantum mechanical selection rules, 
we have $\langle i|
\hat{\bf S}|0\rangle = \langle i|\hat{\bf S}|1\rangle = 0$ for 
$|i\rangle \neq |0\rangle, |1\rangle$, i.e., a uniform magnetic field 
does not cause leakage to states outside the ground state doublet.

For the CNOT gate, one requires a tunable exchange interaction 
$\hat{H}_{\ast}$ 
between one or several spins of neighboring spin cluster qubits I and II. 
For simplicity, we first restrict our attention to an isotropic 
exchange coupling
$\hat{H}_{\ast}=J_{\ast}(t) \hat{\bf s}_{n_c}^{\rm I} \cdot 
\hat{\bf s}_{1}^{\rm II}$
between the outermost spins of clusters I and II, respectively 
[Fig.~\ref{Fig0}(c)]. 
This
exchange interaction will in general not only couple states
within the two--qubit basis $\{|00\rangle, |01\rangle,|10\rangle,
|11\rangle \}$, but will also lead to transitions to excited states (leakage). 
If $J_{\ast}(t)$ changes adiabatically, i.e., on time
scales long compared to $\hbar/\Delta$ and $|J_{\ast} (t)| \ll \Delta$ for all
times $t$, leakage remains small (see below). The action of
$\hat{H}_{\ast}$ can then be described by an
effective Hamiltonian in the two--qubit product basis
\begin{equation}
\hat{H}_{\ast} = J_{\ast z}(t)  \hat{S}^{\rm I}_z   \hat{S}^{\rm II}_z 
+ \frac{J_{\ast \perp}(t)}{2} ( \hat{S}^{{\rm I}+}   \hat{S}^{{\rm II}-}
+ \hat{S}^{{\rm I}-}   \hat{S}^{{\rm II}+})
 , \label{eq:cnot1}
\end{equation}
where~the~roman~numbers~label~the~spin clusters, $J_{\ast z}(t) =
4 J_{\ast}(t) |_{\rm I}\langle 0|\hat{s}_{n_c,z}^{\rm I}|0\rangle_{\rm I}|
|_{\rm II}\langle 0|\hat{s}_{1,z}^{\rm II}|0\rangle_{\rm II}|$,
and $J_{\ast \perp}(t) = 4  J_{\ast}(t) 
|_{\rm I}\langle 1|\hat{s}_{n_c,x}^{\rm I}|0\rangle_{\rm I}|
|_{\rm II}\langle 0|\hat{s}_{1,x}^{\rm II}|1\rangle_{\rm II}|$.
We have shown that the coupling $\hat{H}_{\ast}$ is 
isotropic also
in the two--qubit product basis and 
acts on the states $|0\rangle$
and $|1\rangle$ of neighboring spin chains in the same way as an
isotropic exchange between two single spins. 
$|_{\rm I}\langle 1|\hat{s}_{n_c,x}^{\rm I}|0\rangle_{\rm I}|$ and
$|_{\rm II}\langle 0|\hat{s}_{1,x}^{\rm II}|1\rangle_{\rm II}|$
determine the gate operation time $\tau_{\rm CN}$
of the CNOT gate, $|i\rangle|j\rangle \rightarrow 
|i\rangle |i + j \, {\rm mod} \, 2\rangle$ 
where $i,j = 0,1$. For 
$n_c = 9$, \ldots,  $15$, 
the matrix elements are of order $0.1$, i.e., a factor of $5$ smaller than 
for a single spin $1/2$.

Although we have so far discussed one--qubit gate operations induced from 
spatially uniform magnetic fields, such uniformity may be difficult to 
achieve experimentally. One--qubit gates can be 
performed with spatially varying fields $B_{i,x}$ and $B_{i,z}$ (or
$g$ factors) for which 
$|\langle 1|\sum_{i=1}^{n_c} g_i \mu_B B_{i,x} \hat{s}_{i,x}|0\rangle| \neq 0$
 and
$\langle 0|\sum_{i=1}^{n_c} g_i \mu_B B_{i,z} \hat{s}_{i,z}|0\rangle \neq 0$, 
respectively. Similarly, the analysis
leading to Eq.~(\ref{eq:cnot1}) remains valid for a wide class of
coupling Hamiltonians $\hat{H}_{\ast}$ for which
$\langle 10| \hat{H}_{\ast} |01\rangle \neq 0$. For illustration we 
discuss two examples. First, couplings
between several spins of cluster I to several spins of cluster II, such
as $\hat{H}_{\ast} = J_{\ast}  \sum_{i=1}^{n_c}\hat{\bf s}^{\rm I}_i \cdot 
\hat{\bf s}^{\rm II}_i$, are permitted and even lead to a decrease of
$\tau_{\rm CN}$ because the coupling of several spins in the microscopic
Hamiltonian leads to an increased effective coupling between the
clusters. Second, a modification of the intracluster exchange 
couplings by
$\hat{H}_{\ast}$ due to additional terms such as 
$J_{\ast} \hat{\bf s}^{\rm II}_1 \cdot 
\hat{\bf s}^{\rm II}_2$ does not invalidate
the proposed gate operation scheme. This illustrates the most 
significant advantage 
of the spin cluster qubits over single-spin qubits -- that {\it it is 
sufficient to control magnetic fields and exchange interactions on a 
scale of the spin cluster diameter}. For the linear spin cluster qubit, this
length scale is $n_c$ times larger than the original qubit.

A set of universal quantum gates is necessary but not sufficient 
for the realization of a quantum computer. Rather, additional
requirements must be met, including initialization, decoherence times large
compared to gate operation times, and readout~\cite{divincenzo:95}. 
Initialization can be achieved 
by cooling in a magnetic field $B_z$ to a temperature~\cite{loss:98}
$T \lesssim g \mu_B B_z/k_B < \Delta/k_B$.
Because the state of the spin cluster qubit, $|0\rangle$
or $|1\rangle$, determines the sign of the local magnetization at each site
within the spin chain [Fig.~\ref{Fig0}(b)], readout of the spin cluster qubit 
can be accomplished
by readout of the spins within the cluster~\cite{loss:98,recher:00}. 

An important consideration is the effect of decoherence on spin
cluster qubits. The scaling
of the decoherence rate with system size depends on the 
microscopic decoherence mechanism. For electron spins in quantum dots,
fluctuating fields and nuclear spins have been
identified as dominant sources~\cite{loss:98,burkard:99,khaetskii:02}.
We model~\cite{loss:98} the action of fluctuating magnetic fields by 
$\hat{H}_{\phi}^B = b(t) \hat{S}_z$ where $b(t)$ is Gaussian white 
noise, $\langle b(t) b(0)\rangle
= 2 \pi \gamma^B \delta (t)$. 
Because the magnetic moment $\pm g \mu_B/2$ of the spin cluster qubit
is the same as for a single spin, the decoherence rate~\cite{blum:96} 
$\pi \gamma^B$ is 
independent of $n_c$. In contrast, the decoherence rate 
due to fluctuating fields acting independently on each site increases linearly
with $n_c$.

\begin{figure}
\centerline{\psfig{file=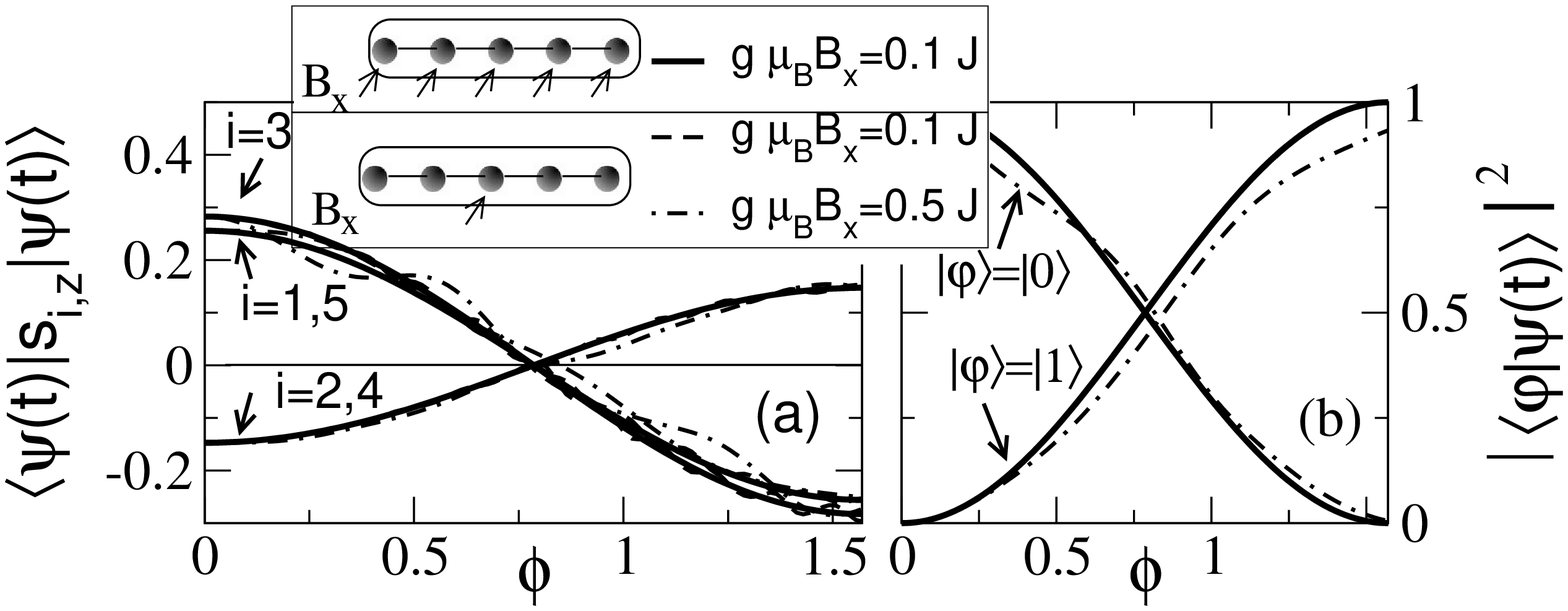,width=8.3cm}}
\caption{
(a) Local spin density 
within a spin cluster qubit ($n_c=5$) as function
of $\phi \propto g \mu_B B_x t/\hbar$ obtained by integration
of the full Schr\"odinger equation for homogeneous (solid line)
 and inhomogeneous $B_x$ (dashed and dashed-dotted lines). 
(b) For $B_x \ll \Delta/g \mu_B$ 
or homogeneous $B_x$, the state rotates coherently from 
$|0\rangle$ to $|1\rangle$. 
For a magnetic field acting only on the central
spin of the cluster, the leakage increases to $7$\% for 
$g \mu_B B_x = 0.5 J$.}
\label{Fig1}
\end{figure}

{\it Spin dynamics during gate operation. --}
One-- and two--qubit gates induce spin 
dynamics in the clusters, and leakage out of the ground state doublet 
is required to remain small. In order to quantify leakage, 
by numerical integration of the Schr\"odinger equation we trace the time 
evolution of a small spin cluster qubit ($n_c=5$) during the 
one--qubit rotation gate. The 
qubit is rotated coherently from $|0\rangle$ into $|1\rangle$, which 
corresponds to a simultaneous rotation of all spins [Figs.~\ref{Fig1}(a)
and (b)]. 
The one--qubit rotation
can also be generated by an inhomogeneous field $B_x$ acting, e.g,  only on 
the 
central spin of the cluster as long as $g \mu_B B_x \ll \Delta$ 
[Figs.~\ref{Fig1}(a) and (b)]. 
Leakage due to
instantaneous switching is less than $0.3$\% for $g \mu_B B_x = 0.1 J$, but
increases with $g \mu_B B_x$ [Fig.~\ref{Fig1}(b)].

\begin{figure}
\centerline{\psfig{file=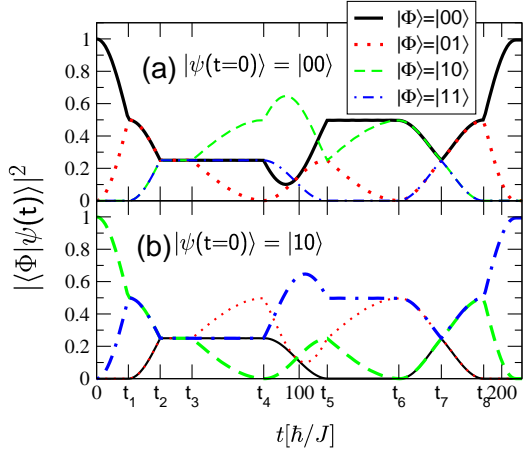,width=7cm}}
\caption{CNOT gate for two small spin cluster qubits ($n_c=3$) obtained by  
numerical integration of the Schr\"odinger equation [see Fig.~\ref{Fig0}(c)].
The plotted probabilities and the
phases (not displayed) show that 
(a) $|00\rangle \rightarrow |00\rangle$ and (b) 
$|10\rangle \rightarrow |11\rangle$. 
We have chosen a pulse
sequence [Eq.~(\ref{eq:cnot2})] with instantaneous switching 
(at times $t_i$), 
$B =0.1 J/g \mu_B$, and
$J_{\ast} = 0.1 J$. Leakage due to instantaneous switching 
($0.7$\% for our parameters) can be 
reduced by decreasing $J_{\ast}$ and $B$.}
\label{Fig2}
\end{figure}

For the special cases  $J_{\ast z}=J_{\ast \perp}$  and 
$J_{\ast z}=0$ 
in Eq.~(\ref{eq:cnot1}), an explicit pulse 
sequence for the CNOT gate has been derived previously in 
Refs.~\cite{loss:98,burkard:99b}. 
We define the unitary time evolution
operator $U_{\ast} (\pi/2) = {\mathrm T} \exp 
\left(-i \int dt \, \hat{H}_{\ast} /\hbar\right)$,
with
$-\int dt \, J_{\ast \perp} (t) /\hbar = \pi/2$.  Then, more generally,
\begin{eqnarray}
U_{\rm CNOT} &\sim&  e^{-i \pi S_{y}^{\rm II}/2}
e^{i 2 \pi {\bf n}_1 \cdot {\bf S}^{\rm I}/3} e^{i 2 \pi {\bf n}_2 
\cdot {\bf S}^{\rm II}/3}
 U_{\ast}(\pi/2)  \nonumber \\ && \hspace*{0.5cm}
\times  e^{i \pi S_{y}^{\rm I}} U_{\ast} (\pi/2)
e^{-i \pi S_{x}^{\rm I}/2} e^{-i \pi S_{x}^{\rm II}/2}  
e^{i \pi S_{y}^{\rm II}/2} \label{eq:cnot2}
\end{eqnarray} 
is the CNOT gate for an arbitrary effective XXZ--coupling Hamiltonian 
[Eq.~(\ref{eq:cnot1})] if
$J_{\ast \perp} \neq 0$,
where  ${\bf n}_1= (1,-1,1)/\sqrt{3}$ and ${\bf n}_2= (1,1,-1)/\sqrt{3}$ .
We confirmed that for the complete pulse sequence 
the dynamics of two spin clusters is as predicted  
on the basis of the two-level description [Fig.~\ref{Fig2}]. Leakage 
induced by $\hat{H}_{\ast}$ is small for $J_\ast \ll \Delta$ because all 
spins in the 
clusters corotate although $\hat{H}_\ast$ couples only the outermost spins.

{\it Spatially varying and anisotropic exchange. --}
We show next that spin cluster qubits are extremely
robust with respect to spatial variation (accounted for by $f_i$ in 
Eq.~(\ref{eq:anchain}))
and anisotropies ($J_\perp \neq J_z$) of the intracluster exchange. 
For spatially varying isotropic exchange
($J_\perp = J_z=J$), the system still exhibits 
a $S=1/2$ ground state doublet~\cite{lieb:62} and the 
above analysis remains valid. In systems 
such as quantum dot arrays where it is possible to engineer the 
intracluster exchange $J f_i$ during sample growth, the qubit basis states 
$\{|0\rangle,|1\rangle \}$ can be tailored to some extent.

We next consider the XY chain, $J_z=0$. By the
Jordan--Wigner transformation~\cite{auerbach:94},
the XY spin chain is mapped onto a system of noninteracting spinless 
fermions  with spatially varying hopping amplitudes,
$\hat{H} = - (J_\perp/2) \sum_{i=1}^{n_c-1} f_i 
(\hat{\psi}_{i+1}^\dagger \hat{\psi}_i +\hat{\psi}_{i}^\dagger 
\hat{\psi}_{i+1} )$,
where $\hat{\psi}_i$ annihilates a Jordan-Wigner fermion at site $i$. We
find that the one--particle Hamiltonian 
has $(n_c-1)/2$ states with negative and positive energy, 
respectively, which are pairwise related to each other by staggering of the 
wave function. There is one zero--energy eigenstate 
\begin{equation}
{\bf e}_0 \propto \left(1,0,-\frac{f_1}{f_2},0,
\frac{f_1 f_3}{f_2 f_4},0, \ldots, \pm  
\frac{f_1 f_3 \ldots f_{n_c-2}}{f_2 f_4 \ldots f_{n_c-1}}\right).
\label{eq:xy-eigenstate}
\end{equation}
The ground state doublet of the XY chain corresponds to the lowest $(n_c-1)/2$ 
and $(n_c+1)/2$ Jordan--Wigner fermion levels filled.
For $f_i \equiv 1$, $\Delta \simeq  \pi J_\perp /n_c$. Similarly to the
spin chain with isotropic exchange, 
one--qubit gates can be realized by magnetic fields $B_z(t)$ and $B_x(t)$ 
unless
$\langle 1|\hat{S}_x|0\rangle =0$. For $n_c\leq 9$ and $f_i \equiv 1$, 
$|\langle 1|\hat{S}_x|0\rangle| \geq 0.4$.
From Eq.~(\ref{eq:xy-eigenstate}), one can also calculate all matrix elements 
entering Eq.~(\ref{eq:cnot1}). In particular, for $f_i \equiv 1$, 
$\langle 0| \hat{s}_{n_c,z}|0\rangle = 1/(n_c+1)$, and 
$|\langle 1| \hat{s}_{n_c,x}|0\rangle| = 1/\sqrt{2(n_c+1)}$.
Because of the anisotropy of the intrachain exchange, $\hat{H}_{\ast}$
(which is isotropic in the single spin operators) translates into an 
anisotropic effective Hamiltonian Eq.~(\ref{eq:cnot1}). Nevertheless, 
the CNOT gate can still be realized according to 
Eq.~(\ref{eq:cnot2}). For the anisotropic chain, a magnetic field applied along
an axis ${\bf n}$ translates into a rotation around the axis 
$\propto ( 2|\langle 1|\hat{S}_x|0\rangle|n_x, 
 2|\langle 1|\hat{S}_x|0\rangle| n_y,n_z)$ in the Hilbert space 
spanned by $\{|0\rangle,|1\rangle\}$. A one--qubit rotation around an
arbitrary axis hence requires appropriate rescaling of ${\bf B}$.
For example, the rotation corresponding to
$\exp(i 2 \pi {\bf n}_1 \cdot {\bf S}^{\rm I}/3)$ [Eq.~(\ref{eq:cnot2})] for 
the
isotropic chain can be achieved by applying a magnetic field 
$B =B_0 (1+2/(2|\langle 1|
\hat{S}_x|0 \rangle|)^2)^{1/2}
/\sqrt{3}$ along the axis $\propto (1,-1,2|\langle 1|\hat{S}_x|0 \rangle|)$ 
for a time $2 \pi \hbar/3 g \mu_B B_0$.
For given $J_{\ast}$ and $B$, the CNOT
gate operation time increases at most linearly with $n_c$.

For $J_z \gg J_\perp$~\mbox{(Ising--like~systems),~where} \mbox{$|0\rangle =
|\uparrow\rangle_1
|\downarrow\rangle_2\ldots |\uparrow\rangle_{n_c}$} 
$+ {\mathcal O}(J_\perp/J_z)$, the ground state doublet 
is~separated~from the next excited state by an $n_c$--independent 
$\Delta \sim J_z \min(f_i)$. In perturbation theory 
in $J_\perp/J_z$, for $f_i \equiv 1$, the matrix elements
$|\langle 1|\hat{S}_x|0\rangle|,  |\langle 1|\hat{s}_{n_c,x}|0\rangle| 
\sim 
(2J_\perp/J_z)^{(n_c-1)/2}$
decrease exponentially with $n_c$. 
Even for medium sized chains $n_c \gtrsim 9$ 
and $J_\perp/J_z < 0.2$, an isotropic $\hat{H}_{\ast}$ translates into an 
effective Ising
Hamiltonian, $J_{\ast \perp}\simeq 0$ in 
Eq.~(\ref{eq:cnot1}). Hence, only quantum 
computing schemes which 
rely on Ising interactions~\cite{raussendorf:01} are feasible. 

{\it Discussion. --} 
The main
idea of the present work applies not only to spin chains but 
remains valid for a wide class of antiferromagnetic
systems with uncompensated sublattices, also in higher dimensions $d>1$ and
for larger spins $s>1/2$. 
We illustrate the advantages of spin cluster qubits for electron spins in 
quantum dots with a typical diameter of $d=50$~nm, where the exchange
coupling can be as large as $10$~K~\cite{burkard:99}. One--qubit operations
are realized, e.g., by $g$--factor engineering in presence of a static field
$B\simeq 1$~T. We now compare the performance of a spin cluster qubit formed 
by 
$n_c=5$ quantum dots coupled by an intracluster exchange $J=10$~K with a 
single
spin qubit. To obtain an estimate, we consider gate operation by
switching the magnetic field $B$ to $g \mu_B B=0.7$~K, and 
$J_\ast = 2.3$~K~\cite{burkard:99},
small compared to the energy gap $\Delta = 7.2$~K of the spin cluster.
For single spins, the gate operation times for the NOT
and CNOT gate are $36$~ps and $117$~ps, respectively. Assuming that 
the magnetic field decreases smoothly from its maximum value at the central 
spin of the spin cluster qubit to $0.2B$ acting on spins $2$ and $4$, we 
find that the operation
time for one--qubit gates increases by a factor $1/2|\langle 1|\hat{s}_{3,x}
+ 0.2 \hat{s}_{2,x}+ 0.2 \hat{s}_{4,x} |0\rangle| = 2.2$ compared to the 
single spin. Similarly, for the operation time of the CNOT gate we find 
$280$~ps for the spin cluster qubit. The main advantage of the spin cluster 
qubit is that it is sufficient to control magnetic fields or $g$ factors
on a length scale of $n_c d= 250$~nm
and exchange couplings on a scale of $2 n_c d = 500$~nm. This would allow 
one to control the exchange between 
neighboring clusters optically~\cite{levy:01} at the expense of an increase
in gate operation times by a factor of $2$.

Other possible applications for spin cluster qubits include, e.g., 
P atoms in a 
Si matrix~\cite{kane:98} and molecular magnetic systems~\cite{gatteschi:94}. 
For electron spins of P atoms  
in a Si matrix, the requirement of positioning P with lattice spacing 
precision~\cite{kane:98,koiller:02} can be circumvented by the use of spin 
clusters instead of single spins. More generally, 
the present work shows that for universal quantum gates 
control is not required at the level of single electron
spins. 
Because a qubit can always
be mapped onto a spin $1/2$, the general principle of arranging several 
qubits into a cluster qubit  applies to any quantum computing 
proposal.

{\it Acknowledgements. --} This work was  
supported by the EU TMR network no. HPRN-CT-1999-00012 
(FM and DL), DARPA SPINS and QuIST 
(JL and DL), the Swiss NCCR Nanoscience (DL), and the Swiss NSF (DL). We
acknowledge discussions with G.~Burkard, V.~Cerletti, H.~Gassmann, 
F.~Marquardt, and P.~Recher.

\end{document}